# Diffraction of a partial temporal coherent beam from a single-slit and a circular aperture

E. Koushki*, S. A. Alavi**

*Department of Physics, Hakim Sabzevari University, Sabzevar, 96179-76487, Iran*

*eh.koshki@hsu.ac.ir
** s.alavi@hsu.ac.ir

**Abstract:**

We generalize the notion of the Franhoufer diffraction from a single slit and a circular aperture to the case of partially temporal coherent and quasimonochromatic light. The problem is studied analytically and the effect of coherence length on the diffraction pattern is investigated. In this case the far-field distribution of the irradiance depends on the newly introduced parameter $\eta$ "decoherence parameter" which governs the deviation of the diffraction pattern from the usual one. The $\eta$-dependent corrections due to temporal decoherency on the irradiance distribution in the far field is obtained. Numerical study of the effect of decoherence parameter on the Far-field diffraction pattern is performed. In the case of a single slit, there is no noticeable deviation in the central peak, but in the higher orders of diffraction, deviation become apparent. For circular apertures, as long as $\eta > 1$, the beam decoherency affects the distribution profile and the first order diffraction pattern decreases and by increasing the decoherence parameter, the first order of diffraction pattern gradually disappears.

# Introduction

Diffraction of partial spatial coherent illumination has been studied in several works. Parrent and Skinner discussed the one-dimensional problem in some detail [1]. They assumed that the quasi-monochromatic illumination across the diffracting aperture was of uniform intensity and they obtained the degree of coherency as a function of ratio of vector positions on the aperture to the coherence interval. Their model takes a general solution for the far-field intensity. Bakos and Kantor used the Young-Rubinovicz theory of diffraction, in which diffraction arises from interference of the incident wave with secondary waves emerging from the diffracting edge to obtain the resultant intensity distribution in the diffraction pattern [2]. They studied the one-dimensional problem both experimentally and theoretically. Cathey discussed diffraction by an odd slit and exponential correlation function was assumed in the integration [3]. Thompson experiments showed that previous models should be refined [4]. Shore, Thompson and Whitney presented other correlation forms for degree of coherence that satisfied experiments [3]. Extensive theoretical and numerical efforts have been made by Shore [5] and Asakura [6], to develop Fraunhofer diffraction with partial spatial coherent light.

On the other hand, multiple-beam interference pattern under partial spatial coherent illumination has been studied in several works. Both one- and two-dimensional regular and irregular arrays of apertures are treated for interference with partial coherent light by Thompson [7]. Investigation of three- and four-beam interference under partial coherent illumination from theoretical and experimental points of views has been carried out by Nawata and Suzuki [8-10]. The general formula for the intensity distribution in the multiple-beam interference pattern under the partial spatial coherent illumination is derived in [11].

Gbur et al. studied the interference of three point sources which are mutually partial coherent and they shown that complete destructive interference can occur even if the fields coming out from the pinholes are not fully coherent with respect to each other [12]. Experimental verification of this model was made by Ambrosini et al. [13].

Optical models of slit diffraction experiments have been developed to particles like electrons and neutrons [14,15] and also x-ray experiments [16,17] to explain the behavior of the empirical data.

In contrast to the diffraction of partial spatial case, the problem of diffraction of partial temporal coherence has not been widely studied in the literature. The interaction of light with different devices may change the time coherence, so it is interesting and also of practical importance to study temporal coherent beam. The laser beam is almost perfect spatial coherent at the laser output, but transferring through optoelectronic, switcher, modulator devices and choppers change the temporal coherence [18-21]. Therefore, in the situations that the temporal coherence is important for the system performance, the effect of change of the temporal coherence time must take into account. For instance, in optical communications, temporal coherence properties of ultra-short pulses are studied to select an appropriate laser structure in order to system performance [22-25]. On the other hand, diffraction phenomena in optical communications have a key role in communication network system. Advantages of diffraction devices and gratings make them good candidates for applications in many cases in optical communications including, complex filtering, pulse shaping, processing and precise chromatic dispersion compensation [26-28]. So, the study of effects of the temporal coherence on diffraction phenomena is unavoidable in optical communications.

In this paper we study Fraunhofer diffraction from a single slit illuminated by a partial temporal coherent beam which is perfect spatial coherent over all the points on the slit. This assumption is satisfactory for an enough thin slit and also for laser beams. A general formula is derived to describe the intensity distribution of the Fresnel diffraction pattern of a slit aperture illuminated with partial temporal coherent light.

**Theoretical analysis**

In Fig.1, Interference at point $P$ due to three waves from the source traveling along different paths is shown. The electric field at point $P$ despite a constant factor $\frac{c\varepsilon_0}{2}$ could be written as follows:

$$\vec{E}_p(t) = \vec{E}_1(t) + \vec{E}_2(t+\tau) + \vec{E}_3(t+2\tau) = \vec{E}_{01}e^{-i\omega t} + \vec{E}_{02}e^{-i\omega(t+\tau)} + \vec{E}_{03}e^{-i\omega(t+2\tau)} \qquad (1)$$

where $\tau$ is the constant time difference (time dilation) between two neighboring paths. The irradiance at $P$ is given by:

$$I_p = <\vec{E}_p.\vec{E}_p^*> = <(\vec{E}_1+\vec{E}_2+\vec{E}_3).(\vec{E}_1^*+\vec{E}_2^*+\vec{E}_3^*)> =$$
$$<|E_1|^2+|E_2|^2+|E_3|^2+(\vec{E}_1.\vec{E}_2^*+\vec{E}_1.\vec{E}_3^*+\vec{E}_2.\vec{E}_3^*+\vec{E}_3.\vec{E}_2^*+\vec{E}_2.\vec{E}_1^*+\vec{E}_3.\vec{E}_1^*)> \qquad (2)$$

which can be written as:

$$I_p = I_1+I_2+I_3+2[\mathrm{Re}<E_1 E_2^*>+\mathrm{Re}<E_2 E_3^*>+\mathrm{Re}<E_1 E_3^*>]=$$
$$I_1+I_2+I_3+2\sqrt{I_1 I_2}\,\mathrm{Re}(\gamma_{12}(\tau)+2\sqrt{I_2 I_3}\,\mathrm{Re}(\gamma_{23}(\tau))+2\sqrt{I_1 I_3}\,\mathrm{Re}(\gamma_{13}(2\tau)) \qquad (3)$$

where the angle brackets represent time averages. If $I_1 = I_2 = I_3 = |E_0|^2 = I_s$, where $I_s$ is the self-product intensity of each individual segment, then the normalized correlation function (the degree of coherence) $\gamma_{ij}(t)$ is:

$$\gamma_{12}(\tau) = \frac{<E(t)E^*(t+\tau)>}{I_s}$$

$$\gamma_{23}(\tau) = \frac{<E(t+\tau)E^*(t+2\tau)>}{I_s}$$

$$\gamma_{13}(\tau) = \frac{<E(t)E^*(t+2\tau)>}{I_s} \qquad (4)$$

Considering the electric field with harmonic frequency $\omega$ and the phase $\varphi(t)$, $E(t)=E_0 e^{-i(\omega t - \varphi(t))}$ the product of the fields would be:

$$E(t+\tau)E^*(t+2\tau) = E_0^2 e^{i\omega\tau} e^{i[\varphi(t+\tau)-\varphi(t+2\tau)]} \qquad (5)$$

so that

$$\gamma_{12}(\tau) = e^{i\omega(\tau)} <e^{i[\varphi(t)-\varphi(t+\tau)]}>$$

$$\gamma_{23}(\tau) = e^{i\omega\tau} <e^{i[\varphi(t+\tau)-\varphi(t+2\tau)]}>. \qquad (6)$$

By the same way:

$$\gamma_{13}(2\tau) = e^{i\omega(2\tau)} <e^{i[\varphi(t)-\varphi(t+2\tau)]}>. \qquad (7)$$

After some calculations [29,30], one could write:

$$\gamma_{12}(\tau) = \gamma_{23}(\tau) = (1-\frac{\tau}{\tau_0})e^{i\omega(\tau)}$$

$$\gamma_{13}(2\tau) = (1 - \frac{2\tau}{\tau_0})e^{i\omega(2\tau)}. \tag{8}$$

where $\tau_0$ is the coherence time. So the Eq.(3), can be written as:

$$I_p = 3I_s + 2I_c[2(1 - \frac{\tau}{\tau_0})\cos(\omega\tau) + (1 - \frac{2\tau}{\tau_0})\cos(2\omega\tau)] \tag{9}$$

where $I_c = \sqrt{I_1 I_2} = \sqrt{I_2 I_3} = \sqrt{I_1 I_3}$ is the cross–product intensity due to the contribution of two mutual intensities.

The first term in the bracket is due to the interference of the waves from two pairs paths (1,2) and (2,3), for which the relative time dilation is $\tau$. The second term refers to the interference of the waves from two paths (1,3) for which the relative time dilation is $2\tau$.

For $N$ paths with the same irradiances $I\ (= I_s = I_c)$ and constant relative time dilation $\tau$ between adjacent paths, we can generalize Eq.(9) as follows (see appendix A):

$$\begin{aligned}I_P &= NI_s + 2I_c[(N-1)(1 - \frac{\tau}{\tau_0})\cos(\omega\tau) + (N-2)(1 - \frac{2\tau}{\tau_0})\cos(2\omega\tau) + \\ &(N-3)(1 - \frac{3\tau}{\tau_0})\cos(3\omega\tau) + ....... + (N - (N-1))(1 - \frac{(N-1)\tau}{\tau_0})\cos((N-1)(\omega\tau)]\end{aligned}. \tag{10}$$

This equation shows that for different $N$ paths with consecutive time dilation $\tau$, there are $N-1$ contribution terms with time dilation $\tau$, $N-2$ contribution terms with time dilation $2\tau$ and finally one contribution terms with time dilation $(N-1)\tau$ which comes from the interference between the first and the last paths.

Eq.(10) may be rewritten as:

$$I_P = NI_s + 2I_c \sum_{j=1}^{N-1} (N-j)(1 - \frac{j|\tau|}{\tau_0})\cos(j\omega\tau). \tag{11}$$

As mentioned earlier, $\tau$ is the relative time dilation which can be both negative or positive but we are dealing with the absolute value of it, so we have used $|\tau|$ in Eq.(11). For $\tau_0 \to \infty$, it gives the diffraction distribution for the perfect coherence case [29]:

$$I_P = NI_s + 2I_c \sum_{j=1}^{N-1} (N-j)\cos(j\omega\tau). \tag{12}$$

This result is appropriate for discrete wave sources. Now we consider the single slit as a continuous series of spot sources. We suppose the slit width is $b$ which contains $N'$ spot sources and is divided into $N$ equal segments as shown in Fig.(2). We define $E_L$ as the amplitude per unit width of slit at unit distance away which is given by [29,30]:

$$E_L = \frac{1}{b} \lim_{N' \to \infty} (\varepsilon_0 N') \tag{13}$$

where $\varepsilon_0$ is the electric field due to each spot source. This definition avoids the irradiation of infinite number of sources to become divergent. The electric field of $j$-th segment with width $\Delta y_j$ at distance $r$ is $\frac{1}{br} \lim_{N' \to \infty} (\varepsilon_0 N') \Delta y_j$. In this case the second term of Eq.(11) changes to the following form:

$$2 \lim_{N' \to \infty} \lim_{N \to \infty} \sum_{j=1}^{N-1} (\frac{1}{br_j}(\varepsilon_0 N')\Delta y_j)^2 (N-j)(1 - \frac{j|\tau|}{\tau_0}) \cos(j\omega\tau). \tag{14}$$

Using Eq.(13) it can be rewritten as;

$$2 \lim_{N \to \infty} \sum_{j=1}^{N-1} (\frac{E_L}{r_j} \Delta y_j)^2 (N-j)(1 - \frac{j|\tau|}{\tau_0}) \cos(j\omega\tau). \tag{15}$$

By using $|\tau| = \frac{b|\sin\theta|}{Nc}$ it can be transformed into the following form:

$$2 \lim_{N \to \infty} \sum_{j=1}^{N-1} (\frac{E_L}{r_j} \Delta y_j)^2 N(1 - \frac{y_j}{b})(1 - \frac{jb|\sin\theta|}{\tau_0 Nc}) \cos(\frac{j\omega b \sin\theta}{Nc}). \tag{16}$$

In the limit of $N \to \infty$, we have, $\frac{b}{N} \to \Delta y_j$, $\frac{jb}{N} \to y_j$, $\frac{j}{N} \to \frac{y_j}{b}$, so:

$$2 \lim_{N \to \infty} \sum_{j=1}^{N-1} (\frac{E_L}{r_j})^2 \Delta y_j \Delta y_j \frac{b}{\Delta y_j}(1 - \frac{y_j}{b})(1 - \frac{y_j|\sin\theta|}{\tau_0 c}) \cos(\frac{\omega y_j \sin\theta}{c}).$$

Therefore Eq.(15) takes the form as follows:

$$2b(\frac{E_L}{r})^2 \int_{y=0}^{y=b} (1 - \frac{y}{b})(1 - \frac{y|\sin\theta|}{\tau_0 c}) \cos(\frac{y\omega \sin\theta}{c}) dy. \tag{17}$$

In Eq.(11), the first term is the summation of $N$ self-product intensities of different segments at the point P, so it is useful to define the intensity per unit width of slit at unit distance away:

$$I_L = \frac{1}{b}\lim_{N'\to\infty}(i_0 N'). \tag{18}$$

Here, $i_0$ is the intensity due to each spot source. The intensity of $j$-th segment with width $\Delta y_j$ at distance r is $I_j = \frac{1}{br^2}\lim_{N'\to\infty}(i_0 N')\Delta y_j$. So the first term of Eq. (11) reads as:

$$I_s = \lim_{N\to\infty}\sum_{j=1}^{N}(\frac{I_L}{r^2}\Delta y_j) = (\frac{I_L}{r^2})\int_{y=0}^{y=b}dy = b(\frac{I_L}{r^2}). \tag{19}$$

Employing Eqs.(17) and (19), the total intensity distribution from a single-slit is as follows:

$$I_P(\theta) = b(\frac{E_L}{r})^2 + 2b(\frac{E_L}{r})^2\int_{y=0}^{y=b}(1-\frac{y}{b})(1-\frac{y|\sin\theta|}{b}\eta)\cos(\frac{y\omega\sin\theta}{c})dy \tag{20}$$

where the dimensionless parameter $\eta$ which we name it "Slit decoherence parameter" is defined as:

$$\eta = \frac{b}{l_0}, \text{ and } l_0 = c\tau_0 \text{ is the length of finite wave train.}$$

Eq. (20) may be written as:

$$I_P(\theta) = I_P^{\eta=0}(\theta) + \Delta I_P^{\eta}(\theta) \tag{21}$$

where

$$I_P^{\eta=0}(\theta) = b(\frac{E_L}{r})^2 + 2b(\frac{E_L}{r})^2\int_{y=0}^{y=b}(1-\frac{y}{b})\cos(\frac{y\omega\sin\theta}{c})dy \tag{22}$$

is the single-slit diffraction intensity of perfect coherent beam and

$$\Delta I_P^{\eta}(\theta) = -2\eta(\frac{E_L}{r})^2\int_{y=0}^{y=b}y|\sin\theta|(1-\frac{y}{b})\cos(\frac{y\omega\sin\theta}{c})dy \tag{23}$$

is the correction on single-slit intensity distribution due to temporal decoherence diffraction. In the case of perfect coherence ($\eta \to 0$) the correction (23) vanishes and Eq. (20) gives the same results as the Fraunhofer diffraction theory which indicates that the far-field diffraction form a single slit is given by $I(\theta) = I_0 (\frac{\sin \beta}{\beta})^2$ where $\beta = \frac{\pi b \sin \theta}{\lambda}$ [30].

The intensity distribution as given by Eq.(20) is plotted in figures (3) - (5) for different values of $\eta$. In our numerical study the wavelength of the partial coherent beam is 632.8nm (He-Ne laser beam). The slit widths in Figures (3), (4) and (5) are 5, 10 and $20 microns$, respectively. As it is observed, by increasing the slit width, the diffraction pattern is shrinking as expected from the usual Fraunhofer diffraction theory. The deviation of the pattern from the case of perfect coherence is negligible for small values of $\eta$. The central peak (zeroth order of diffraction) is almost the same for different values of $\eta$ which means that the zeroth order of diffraction is not highly sensitive to temporal coherence. But obvious difference is seen between perfect coherence and temporal coherence cases in the first and second order of diffractions. For $\eta \approx 1$ clear destructive interference in the diffraction pattern appears which has been also observed in the case of partial spatial decoherency [12,13]. By increasing $\eta$, reversed interference pattern occurs and the higher orders of diffraction shift to higher angles or in other words, the diffraction pattern is broadened.

Here, we extend the theory to the case of a circular aperture which is of great practical importance since most optical elements have circular shape. Consider a circular aperture with the radius $r_a$ is placed at the distance $z$ from the waist of a Gaussian laser beam. The electric field of a Gaussian beam could be written as [31]:

$$E(z,r) = E_0 \frac{w_0}{w(z)} \exp\{-i(kz - \arctan(z/z_0))\} \cdot \exp\{-r^2(\frac{1}{w^2(z)} + \frac{ik}{2R(z)})\} \quad (24)$$

where $r$, $w_0$ and $k$ are the radial coordinate, the beam waist radius, and the wave number, respectively. Also, $w(z) = w_0 \sqrt{1 + (z/z_0)^2}$, $z_0 = \frac{k w_0^2}{2}$ and $R(z) = z(1 + \frac{z_0^2}{z^2})$ are the beam radius at $z$, the diffraction length of the beam and the radius of the wave front at $z$, respectively.

The diffraction pattern in the far-field (at the distance $D$), in the case of temporal coherence beam is given by:

$$E(\rho) = \frac{1}{i\lambda D} \int_{r=0}^{r=r_a} J_0(k\theta r) E(r,z)(1 - \frac{r|\sin\theta|}{r_a}\eta) 2\pi r dr \tag{25}$$

where $\eta$ is the "aperture decoherence parameter" and is defined as $r_a/l_0$. Also, $J_0(x)$ and $\rho$ are the zero-order Bessel function of the first kind and the radial coordinate in the far-field observation plane, respectively. It is worth mentioning that in the limit $\eta \to 0$, Eq.(25) reduces to the formula for the Franhoufer diffraction pattern from a circular aperture [32]. In the paraxial approximation, the distance from the aperture surface to the far-field observation plane ($D$) is related to the radial coordinate and the diffraction angle by $\rho = D\theta$.

In order to do numerical simulation, we consider a He-Ne laser beam ($\lambda = 632.8 nm$) with the beam waist radius of 1mm and the output power of 5mW incident to a circular aperture with the radius of $r_a = 2\mu m$ placed at z=0.1cm. The output beam is became temporally decoherent using a modulator or chopper with any arbitrary coherence time $\tau_0$. Fig.(6) shows the intensity versus the radial coordinate on the far-field screen located at 3m far from the aperture. Curves have been plotted for different values of $\tau_0$ and $\eta$. As shown, for $\eta < 1$ ($r_a < l_0$), no deviation is observed in comparison with the perfect coherency ($\eta \to 0$). But, for $\eta > 1$, the beam decoherence affects the distribution profile and the first order diffraction patters decreases. In Fig.(7) the diffraction patterns are simulated. As it is observed, by increasing the decoherence parameter, the first order of diffraction patterns gradually disappears.

## Conclusion

Temporal coherence is one of the most important issues in modern optics. In contrast to the partial spatial case, the problem of diffraction of partial temporal coherence has not been widely studied in the literature. There are several reasons to study temporal coherent beam, because interaction with different devices may change the perfect time coherence. For instance, it is well-known that in modern laser sciences and optical communications, the laser beam is almost perfect spatial coherent at the laser output, but transferring through optoelectronic and modulator devices change temporal coherency. In this paper we have studied the diffraction of a partial temporal coherent beam from a single-slit. A general formula has been derived for the intensity distribution of the Fresnel diffraction pattern of a slit aperture illuminated with partial temporal coherent light. The corrections due to temporal coherency on the irradiance distribution in the far field are obtained. We introduced the "slit decoherence parameter" $\eta$ which governs the deviation of diffraction pattern from the usual one (perfect coherent beam). Numerical study of the effect of the slit decoherence parameter on the Far-field diffraction pattern is performed and the range in which the role of parameter $\eta$ is important is discussed. There is no noticeable deviation in the central peak, but in the higher orders of diffraction, deviation become apparent. For the case of diffraction from a circular aperture, analytical and numerical analyses have been done and the effect of aperture decoherence parameter was studied. It was found that by increasing the decoherence parameter, the first order of diffraction patterns disappears. The calculations can be generalized to optical grating technology and X-ray diffraction measurements.

## Appendix A:

We use mathematical induction to prove Eq.(11). We assume that Eq.(11) is valid for $N$ spot sources, then we prove that it holds for $N+1$. The irradiance for $N+1$ paths is:

$$I_{N+1} = (N+1)I + 2I \sum_{j=1}^{N} (N+1-j)(1-\frac{j\tau}{\tau_0})\cos(j\omega\tau)$$

We separate the contribution of $N$ paths, so:

$$I_{N+1} = \left\{ NI + 2I \sum_{j=1}^{N-1} (N-j)(1-\frac{j\tau}{\tau_0})\cos(j\omega\tau) \right\} + \left\{ I + 2I \sum_{j=1}^{N} (1-\frac{j\tau}{\tau_0})\cos(j\omega\tau) \right\}$$

The over index in the second term of the first brocket has been changed from $N$ to $N-1$ because its contribution vanishes for $j=N$.

The first bracket is the irradiance of the $N$ paths while the second is the contribution of $(N+1)$th path. The contribution of the $(N+1)$th path include two terms. The first is the direct irradiance of the $(N+1)$th path and the second is the due to the interference of $(N+1)$th path with all other individual paths. Thus one could write:

$$I_{N+1} = I_N + I'$$

where $I'$ is the total contribution of the ($N+1$)th path. This mathematically deduced equation is completely correct and expected from physical point of view, so Eq.(11) is proved using mathematical induction.

**Figures:**

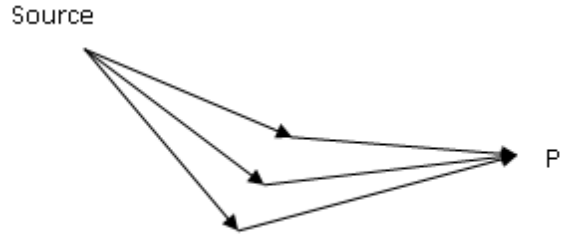

**Fig.1.** Interference of three waves from a source

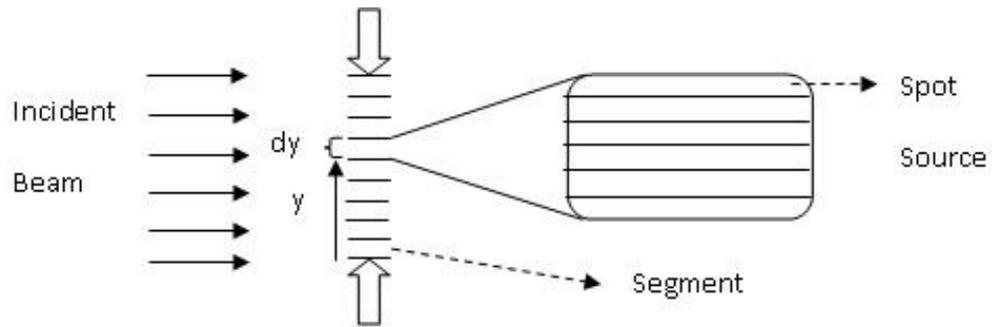

**Fig.2.** Schematically description of a single slit divided **to** $N'$ spot sources and $N$ segments.

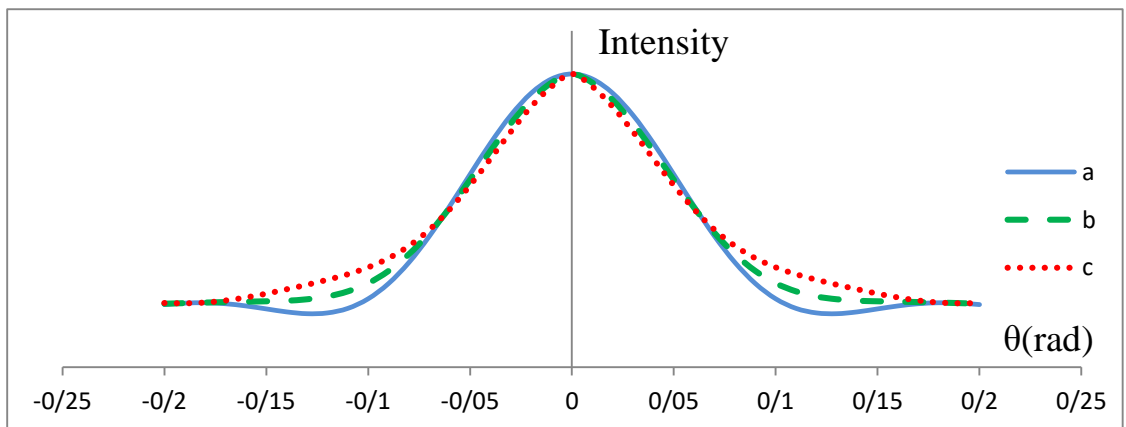

**Fig.3.** Intensity distribution versus diffraction angle for $b = 5 micron$ and for different values of η:

a) η=0.0005, 0.005, 0.05, 0.1, 0.5,1  b) η=5   c) η=10

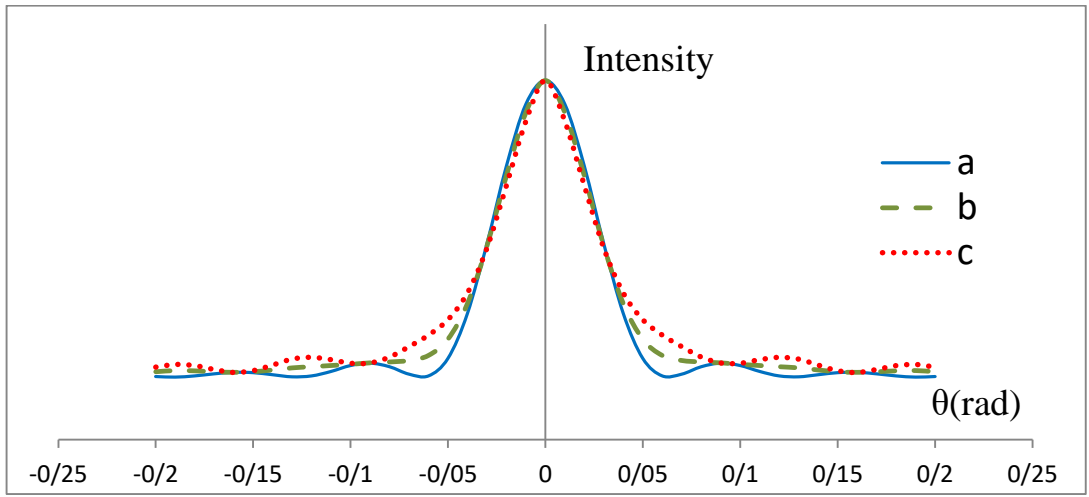

**Fig.4**. Intensity distribution versus diffraction angle for $b = 10 micron$ :

a) η=0.001, 0.01, 0.1, 0.2, 1   b) η=10    c) η=20

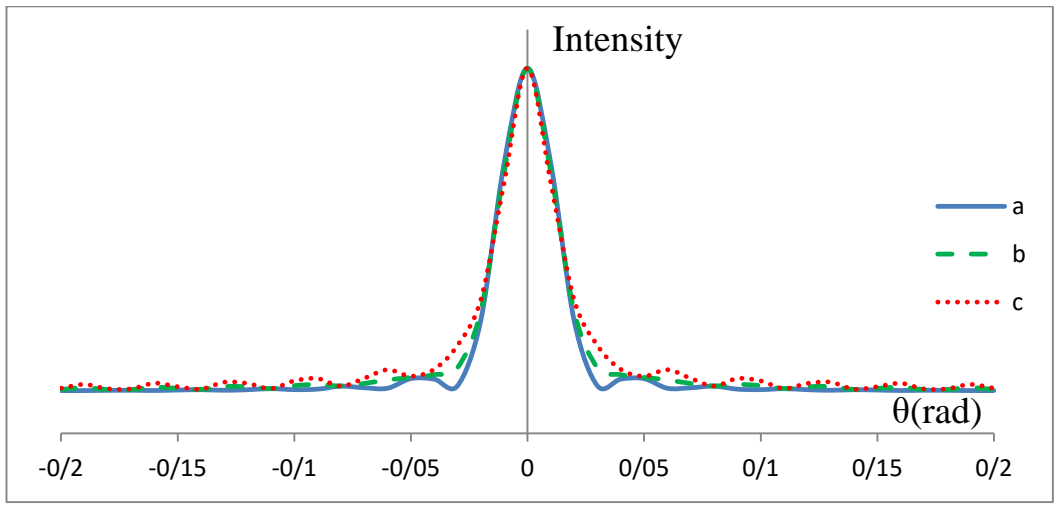

**Fig.5.** Intensity distribution versus diffraction angle for $b = 20 micron$ and for different values of η:

a) η=0.002, 0.02, 0.2, 2, 4   b) η=20    c) η=200

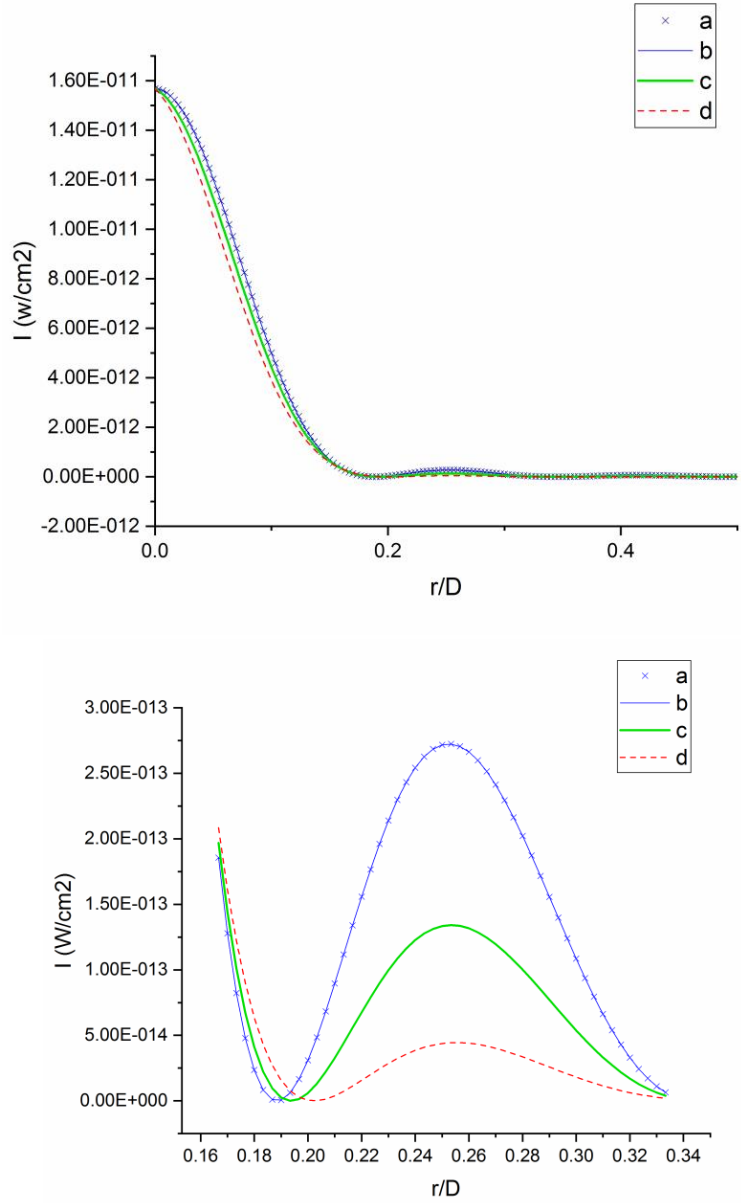

**Fig.6**. Diffraction pattern of a He-Ne Laser beam incident to a circular aperture. The below curve is the magnification on the first order diffraction.

- a- The case of complete temporal coherency $\eta \to 0$.
- b- The case with $l_0 = 10 cm$ ($\eta = 2 \times 10^{-5}$). No difference is appeared in comparison with the case (a)
- c- The case with $l_0 = 2 \mu m$ ($\eta = 1$). First order diffraction begins to disappear.
- d- The case with $l_0 = 1 \mu m$ ($\eta = 2$). The first order diffraction is disappeared.

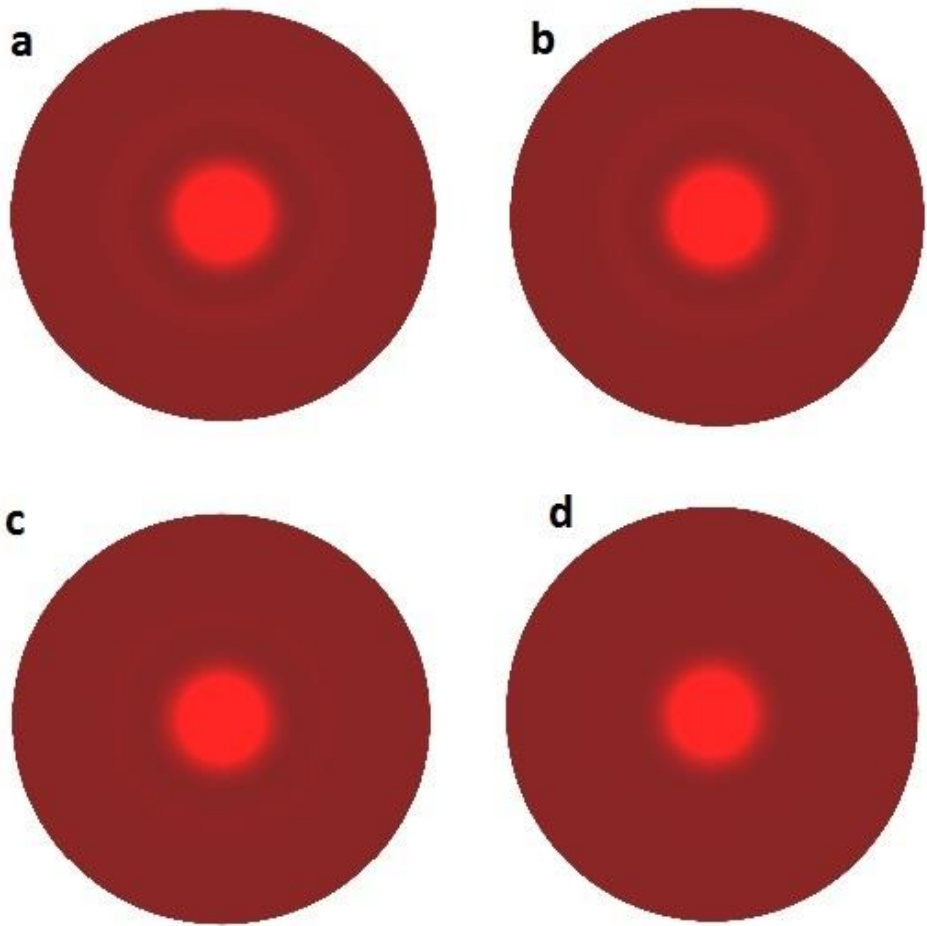

**Fig.7**. Simulation of diffraction pattern of a He-Ne Laser beam incident to a circular aperture:

a- Complete temporal coherency $\eta \to 0$.
b- $\eta = 2 \times 10^{-5}$. No difference is appeared in comparison with the case (a)
c- $\eta = 1$. The light halo (first order diffraction) begins to disappear.
d- $\eta = 2$. The light halo (first order diffraction) is disappeared.